\documentclass{article}
\usepackage[utf8]{inputenc}
\usepackage{gensymb}
\pdfoutput=1

\title{Field free switching through bulk spin-orbit torque in L10-FePt films deposited on vicinal substrates }

\author{Yongming Luo1*,Yanshan Zhuang1,Zhongshu Feng1,Haodong Fan1,Birui Wu1,Menghao Jing1, Ziji Shao1,Hai Li1,Ru Bai1, Yizheng Wu2, Ningning Wang1 and Tiejun Zhou1*}
\date{March 2022}

\begin{document}

\maketitle

\section{Abstract}
L10-FePt distinguishes itself for its ultrahigh perpendicular magnetic anisotropy (PMA), which enables memory cells with sufficient thermal stability to scale down to 3 nm. The recently discovered “bulk” spin-orbit torques in L10-FePt provide an efficient and scalable way to manipulate the L10-FePt magnetization. However, the existence of external field during the switching limits its practical application, and therefore field-free switching of the L10-FePt is in highly demand. In this manuscript, we demonstrate the field-free switching of the L10-FePt by growing it on vicinal MgO (001) substrates. This method is different from previously established strategies, as it does not need to add other functional layers or create asymmetry in the film structure. We demonstrate the field-free switching is robust and can withstand strong field disturbance up to ~1 kOe. The dependence on vicinal angle, film thickness, and growth temperature demonstrated a wide operation window for the field-free switching of the L10-FePt. We confirmed that the physical origin of the field-free switching is the vicinal surface-induced the tilted anisotropy of L10-FePt. We quantitatively characterize the spin-orbit torques in the L10-FePt films, and found the spin-orbit torques are not significantly influenced by the lattice strain from vicinal substrates. Our results extend beyond the established strategies to realize field-free switching, and potentially could be applied to other magnetic and antiferromagnetic systems. 

\section{Introduction}
The FePt in L10 phase possesses one of the highest perpendicular magnetocrystalline anisotropy among transition metal compounds. Its ultrahigh thermal stability enables devices to scale down to 3 nm and distinguishes itself for vast applications, such as high density magntic storages, hard disk drives, magnetic random access memory, and spin logic circuits. [1] However, the large anisotropy also makes the switching of L10-FePt magnetization extremely challenging. Recent experiments have demonstrated that the L10-FePt could be switched by “bulk” like spin-orbit torques (SOT) that exist in the single L10-FePt layer, which provide an efficient and scalable way to electrically manipulate the L10-FePt magnetization. [2-5] However, the deterministic switching of L10-FePt by SOT require an external field colinear with the current [6-8], which increase the complexity of the device structure and operation. [9] Field-free switching of the single L10-FePt film is highly demand for practical applications. In recent years, although various schemes have been developed to realize field-free switching of the perpendicular magnetization, the Field-free switching of the L10-FePt film still has not been achieved yet. [10-14] 

Previously studies have shown that in a magnetic system with tilted perpendicular anisotropy, one could achieve field-free switching. [15] Experimentally, You. et. al first showed that by using the shape induced-tilted anisotropy at the edge of a magnetic pillar, one could realize field-free switching. [16] However, the shape-indued titled anisotropy needs complex etching technique. Liu. et. al further found the tilted anisotropy in the SrRuO3/SrIrO3 bilayer system, and demonstrated the field-free switching in such system. [17] Unfortunately, the SrRuO3/SrIrO3 system could only operate at low temperature, due to its low Curie temperature. Thus, although previously results have demonstrated that a magnetic system with tilted anisotropy could induce field-free switching, it does not attract much attention due to the limited material selection. Currently it is still a critical problem to find a system with tilted anisotropy, which is easy to obtain and suitable for room temperature operation. 
Vicinal substrates denote those substrates with periodic atomic steps exist at the substrate surfaces. They are fabricated by inducing an inclination angle relative to the crystallographic plane during the cutting. The direction and density of atomic steps could be controlled by the inclination angle. Previously, it has been demonstrated that the periodic atomic steps on vicinal substrates could providing an unique way to tune the electronic or chemical properties of the films that are deposited on them, which has been a long-standing subject of intensive research.[18-19] It is established that vicinal substrates could significantly influence the magnetic anisotropy of various ferromagnetic [20-22] and antiferromagnetic films. [23, 24] 

In this manuscript, we demonstrate an artificial way to title the anisotropy of L10-FePt¬by growing it on vicinal MgO (001) substrates, and field-free switching of the single L10-FePt layer could therefore be achieved. We demonstrate the field-free switching is robust and able to withstand strong field disturbance, due to an effective field of ~1kOe induced by the vicinal substrates. Combined with AMR measurements and simulation, we demonstrated that the micro-origin for the realization of field-free switching of L10-FePt is due to the tilting of the easy axis in L10-FePt, when grown on vicinal substrates. We systematically characterize the dependence of the effective field on the vicinal angle of the substrates, the growth temperature and thickness of the L10-FePt film, the results indicate the effective field is proportional to the interface strain of the film. We also quantitively characterize the spin orbit fields, we found the spin orbit fields have similar “bulk” origin as that grown on flat substrates. The amplitude of the SOT is not significantly influenced by the additional strain from the vicinal surfaces.

\section{Result and Discussion}
\subsection{Film characterization:}
L10-FePt is a face-centered tetragonal magnetic alloy, with Fe and Pt atoms alternatively stacked along the c axis, as schematically shown in Figure. 1a. We grow our L10-FePt films by co-sputtering of Fe and Pt on flat or vicinal MgO (001) substrates at high temperatures (700°C if not specified). The vicinal MgO (001) substrates are commercially obtained. Figure. 1b schematically shows the period atomic steps of the vicinal substrate and the films grown on it. \(\alpha\) denotes the vicinal angle of the substrates (shown in Figure. 2b). The L10 phase and perpendicular anisotropy of the FePt films are confirmed by x-ray-diffraction (XRD) patterns and magnetic hysteresis loops, respectively. Figure.1c shows the XRD spectrums of 6 nm FePt films grown on flat (\(\alpha\)=0°) and vicinal MgO (001) substrates (\(\alpha\)=7°). For film grown on the flat substrate, strong FePt (001) and (002) peak could be observed (black line in Figure.1c), demonstrating the high-quality of the L10 -FePt film; while for the film grown on vicinal substrate, although the intensity of XRD spectrum (including the substrate peak) is drastically reduced by three orders compared with that grow on the flat substrate, the (001) FePt peak could still be observed. This demonstrated that the FePt film is still in the L10 phase, even though the chemical order can be reduced compared with that grown on flat substrates.[25] The reduced XRD intensity may due to the larger surface roughness of the vicinal substrates. The magnetic properties of the FePt films are characterized by anomalous Hall measurements. Figure.1d shows the anomalous Hall voltages of the FePt films growing on the flat and vicinal substrates. For both of two samples, the Anomalous Hall signal exhibits square loops with large coercivity (~3 kOe) when the field is swept along the out-of-plane direction, which is consistent with previously reported L10-FePt films.[26] Meanwhile, we could also find there exist distinguishable differences for two different samples, indicating the influence of the vicinal substrates on the magnetic properties of the L10-FePt films. 
We found the easy axis of L10-FePt film would be slightly tilted away from the out-of-plane direction when grown on a vicinal substrate, which can be confirmed by Anisotropy Magnetoresistance (AMR) measurement. If we assume the current is applied in the x-direction, the AMR (\(R_{xx}\)) can be written as:
\[  R_{xx}(\theta )\sim R_{\perp }+(R_{\parallel }-R_{\perp })cos^{2}\phi (\theta )=R_{\perp }+\Delta Rcos^{2}\phi(\theta )                                               (1)\]
where  , \(\theta\) are the angles of FePt magnetization and field direction relative to the x-axis, respectively.   is resistance difference when the magnetization is perpendicular and parallel with the current direction, respectively. We measured the angular dependence of the AMR signal for samples grown on flat (sample \(\#1\), \(\alpha\)=0°) and vicinal (sample \(\#2\), \(\alpha\)=7°) substrates. The set-up of the AMR measurements are schematically shown in Figure. 2a. The current is applied along the vicinal ditrection (y-axis) with a amplitude fixed at 50 uA. Field with different amplitudes are rotated in the yz-plane from 0° to 360° (counterclockwise, CCW), and then rotated back from 360° to 0° (clockwise, CW). The results are shown in Figure.2c. We found both samples exhibit rotational hysteresis around the hard axis (perpendicular with easy axis, \(\theta\) =180°), but have different rotational symmetries. For sample\(\#1\), the hysteresis is symmetric about (\(\theta\) =180°); while for sample \(\#2\), the hysteresis is asymmetric about \(\theta\) =180°, with the center of the hysteresis is also shift from \(\theta\) =180° about 10°. The peak AMR values are also different when the fields rotate CW and CCW. For both samples, the hysteresis loop become smaller and even disappear at a higher field (3 T). Eventhough, we could still find their symmetry axis do not change, the symmetry axis of sample \(\#1\) (sample \(\#2\)) is loacted at \(\theta\) =180° (offset from \(\theta\) =180°) .
Next, by using microspin model and simulation, we demonstrated the asymmetric rotational hysteresis in sample \(\#2 \)is origin from the titled anisotropy in such sysyem. We consider a system with uniaxial anisotropy, according to the microspin model, when the field rotate from easy axis to hard axis (perpendicular with the easy axis), the magnetization would be lagged behind the field, which causes the rotational hysteresis when the field rotate across the hard axis CCW and CW. Besides, it should also be noted that when the magnetization rotate across the hard axis CCW and CW, the switching of the magnetization should be symmetric with it. We consider two situations: (1). Perpendicular anisotropy, with easy axis along z-axis (situation \(\#1\), see the left panel of Figure. 2b); (2) Tilted anisotropy, with the easy axis tilted toward x-axis (situation \(\#2\), see the right panel of Figure. 2b), the tilting angle could be characterized by   (shown in the right panel of Figure. 2b.). For situation \(\#1\), as the hard axis is colinear with the x-axis, the magnetization should be symmetric with x-axis when the magnetization rotate across it CW and CCW, as schematically shown in the left panel of Figure. 2b. The symmetric switching of the magnetization about x-axis also lead to symmetric AMR hysteresis at \(\theta\) =180°. This is because for such AMR measurement geometry, equation (1) denotes the AMR signals is related with the magnetization component in the x-direction (~\({m_{x}^{2}}\) ). While for situation \(\#2\), as the hard axis is not colinear with the x-axis, the magnetization switching should be asymmetric with x-axis (see the right panel of Figure.2b), which would result in asymmetric AMR hysteresis around \(\theta\) =180°. These features could be confirmed by micromagnetic simulation. We simulated the rotation the FePt magnetization when fields with different amplitudes are rotated in the xz plane CW and CCW. The simulation details are shown in supplementary information. We plot the angular dependence of the   at different fields (1.1 T and 2 T) for two systems with perpendicular (\(\varphi =0^{^{\circ} }\)) and tilted (\(\varphi =8^{^{\circ} }\) ) anisotropy. The results are shown in Figure. 2d, at low field (1.1 T), the rotational hysteresis of   is symmetric (asymmetric) with respect to \(\theta\)=180° for situation\(\#1\) (situation \(\#2\)). At high field (2 T), although the hysteresis disappears for both situations due to the large field, the symmetry axis for situation \(\#1\) and \(\#2\) do not change. This results also tell us that the titled anisotropy could be directly confirmed by AMR measurement. Compared with our AMR results (Figure. 2c), we could verify the hysteresis symmetry of sample \(\#1\) (sample \(\#2\)) is well consistence with the situation \(\#1\) (situation \(\#2\)). The consistence between the experiment and theoretic analysis clearly demonstrate that the easy axis of FePt is tilted away from z-axis when grown on vicinal substrates, but fully perpendicular when grown on flat substrate. From the shift of the hysteresis, we can further estimate that the its easy axis is titled about 10° along the vicinal direction for 6 nm FePt grown on vicinal substrate (\(\alpha\)=7°). 
\subsection{ SOT switching characterization: }
Next, we further demonstrate that the tilted anisotropy induced by vicinal substrateis critical to realize field-free SOT switching of the L10-FePt film. To electrically characterize the SOT switching of L10-FePt magnetization, we pattern the FePt films in to Hall bar structures with 5 um in width and 20 um in length, by using photolithography and argon ion milling. To drive the SOT switching of the L10-FePt magnetization, A DC pulse current \({I_{pluse}}\)  with a fixed duration of 30 us along x direction are applied. After each pulse, the magnetization states of FePt film are read out by anomalous Hall resistance (\(R_{H}=V_{ac}/I_{ac}\) ), by applying a small AC current ( \(I_{ac}=50\mu A\)). To help the switching, external fields with different amplitudes \(H_{x}\)  are applied along the current direction, as schematically shown in Figure.3a. We compare the SOT switching behavior for films grown on flat and vicinal substrates. For films grown on flat substrates, we could observe typical SOT-induced magnetization switching, which is the same as what reported previously.[2, 3] We observed the reversal of switching polarity when \(H_{x}\)  changes from positive to negative, and no switching at zero field (see in the supplementary information). While for FePt film grown on vicinal substrate, as shown in Figure.2b, we could also observe the SOT induced switching of magnetization. Remarkably, we found the SOT induced switching can happen even at zero field, demonstrating the realization of field-free switching of the L10-FePt film, which has not been achieved in previously studies. The switching polarity does not change until  . This indicates that the field-free switching of L10-FePt could withstand a strong field disturbance (~1.2 kOe), the vicinal substrate could induce an in-plane effective field ( \(H_{\mathit{eff}}\) ) of 1.2 kOe. Comparing the switching loop with the anomalous Hall loop, we found a partial switching behavior, and the switching ratio is about \(10\%\), this is similar with the previously reported SOT switching of the single L10-FePt film, and may be result from the thermal effects.[2, 3] 
It should also be noted that we found the field-free switching is could not be realized when the pulse current is applied perpendicular to the vicinal direction (along y-axis) (shown in the supplementary information). This feature demonstrate that the field-free SOT switching could only be observed when the current is perpendicular to titled direction of the easy axis of the L10-FePt film (y-axis, as demonstrated in Figure. 2), but could not when they are parallel. Such switching property could be re-produced by micromagnetic simulation (shown in the supplementary information). The consistent of experiment results and simulation clearly demonstrated the physical origin of field-free switching of the L10-FePt film is the titled anisotropy. It should also be noted such switching geometry is also in consistent with the previously reported field-free switching of the SrRuO3/SrIrO3 system.[17]
Besides of substrates with vicinal angle \(\alpha\)= 7°, the field-free switching could also be achieved when FePt films are grown on substrates with other different vicinal angles (\(\alpha\)= 5°, 10°). Figure. 3c plot the change of SOT Hall resistance \(\Delta R_{H}\) (maked in Figure. 3b) as a function of \(H_{x}\) , for films grown on substrates with different vicinal angles (\(\alpha\)=0°, 5°, 7°, 10°). The “±” sign of \(\Delta R_{H}\) in Figure. 3c denote CCW (“+”) and CW (“-”) switching polarity, while \(\Delta R_{H}=0\)  denoting no switching could happen. For the for film grown on flat substrates (\(\alpha\)=0°), we found the \(\Delta R_{H}\) is an odd function of   with \(\Delta R_{H}=0\)  at zero field, this demonstrated field-free switching could not happen. In addition, the result also shown that \(H_{x}\) less than 300 Oe is difficult to induce observable change of \(\Delta R_{H}\). This may be the reason that field-free switching of L10-FePt is relative difficult to realize. By other established schemes, such as by adding an antiferromagnetic layer, or by using wedge shaped film structure, the effective field is usually not large enough (typically about 50~100 Oe).[10, 27, 12] For films grown on vicinal substrates, we have \(\Delta R_{H}\neq 0\) at zero field, indicating the realization of field-free switching. The typical point where \(\Delta R_{H}=0\) (\(H_{x}|_{\Delta R_{\mathit{H}}=0} \)) reflect the effective field created by the vicinal substrate, which can be defined as  \(H_{\mathit{eff}}\)  (marked in Figure.3c). For films grown on different vicinal substrates, we plot  \(H_{\mathit{eff}}\)  as a function of \(\alpha\), as shown in Figure.3d. We found  \(H_{\mathit{eff}}\)   first increase with \(\alpha\)(from 0° to 7°), and then saturates at 7°, as  \(H_{\mathit{eff}}\) at 7° and 10° are very close. Besides, we also studied the switching behavior for FePt layer with different film thickness, as shown in Figure. 3e, we found   \(H_{\mathit{eff}}\)  decrease with film thickness, the effective field decrease from ~1 kOe to 0.5 kOe when the FePt thickness increases from 6 nm to 12 nm, this indicate that the interface origin of   \(H_{\mathit{eff}}\). The dependence of  \(H_{\mathit{eff}}\)  on vicinal angle (\(\alpha\)) and film thickness clearly demonstrated that  \(H_{\mathit{eff}}\)  is physically origin from the interface strain from the vicinal substrates. In addition, we further explored the tunability of  \(H_{\mathit{eff}}\)  by controlling the chemical order of the FePt film. As the XRD intensity of FePt film grow on vicinal substrates is very low, we could not directly obtain the chemical order of the film. We alternatively study the dependence of   \(H_{\mathit{eff}}\)  on the growth temperature, as it is established that the chemical order of FePt increase with the growth temperature.[25] As shown in Figure. 3f, we found  \(H_{\mathit{eff}}\)  increase with annealing temperature, this indicate that the effective field is proportional to the chemical order, which could be attribute to higher lattice strain in films with high crystal order.

\subsection{Second harmonic measurement: }
Next, we quantitively characterize the current induced SOT field in L10-FePt films by harmonic Hall voltage analysis.[28,29] We apply a small AC excitation current (with a density of \(J_{e}\) ) and measure the first (\(V_{\omega }\)) and second (\(V_{2\omega }\)) harmonic signals. Figure. 4a shows the schematic drawing of the measurement set-up and the SOT effective fields   \(\Delta H_{L}\)and \(\Delta H_{T}\) . Figures.4b-d plot the field dependence of \(V_{\omega }\)  and \(V_{2\omega }\)  signals for the 6 nm L10-FePt grown on the vicinal substrate (\(\alpha\)=7°). We obtain the longitudinal (L) and transverse (T) spin orbit fields by: \(\Delta H_{L(T)}=-2\frac{B_{L(T)}\pm 2\xi B_{T(L)}}{1-4\xi ^{2}}\) , where  is defended as \(\left (\frac{\partial V_{2\omega }}{\partial H}/\frac{\partial^2 V_{2\omega }}{\partial H^2}  \right )_{H_{L(T)}}\)   .\(\xi \) is the ratio of the planar Hall voltage to the anomalous Hall voltage. The \(\pm \)  sign corresponds to magnetization pointing along the  \(\pm z\)-axis (i.e.,  ), respectively. [28, 29] Previously studies have demonstrated that in the L10-FePt system although the thermoelectrical effects would influence the second harmonic signal  \(V_{2\omega }\), it only cause an constant offset of the  \(V_{2\omega }\), but has negligible influence on the value of \(\frac{\partial H_{2\omega }}{\partial H}\) , and thus can be neglected. [3] Our results show that \(\Delta H_{L(T)}\)  increases linearly with increasing current density \(J_{e}\), the symmetry of    \(\Delta H_{L}\)( \(\Delta H_{T}\) ) is odd (even) with current  \(J_{e}\) (See in the supplementary information), these features are the same as previously reported current induced effective fields in L10-FePt.[2, 3] To compare the SOT with previous studies, we further calculate the SOT efficiency \(\beta _{L(T)}\), which is defined by \(\Delta H_{L(T)}/\beta _{L(T)}\) . We characterize  \(\beta _{L(T)}\) for different FePt thicknesses, as shown in Figure.4e. We found that both  \(\beta _{L}\) and \(\beta _{T}\)  monotonically increase with the film thickness, demonstrating the “bulk” nature of the SOT. This result is self-consistence with decreasing critical SOT switching current density with the L10-FePt film thickness (see in the supplementary information). The amplitudes of \(\beta _{L(T)}\)  are comparable with previous results.[2, 3] These features demonstrate that the origin of SOT for our FePt films is the same as that grown on flat substrates. We also characterize the dependence of  \(\beta _{L(T)}\) on the growth temperature, which reflected the relation between \(\beta _{L(T)}\)  and the chemical order of the film. As shown in Figure. 4f, we found that \(\beta _{L}\)  and \(\beta _{T}\)  have opposite temperature dependence, with \(\beta _{L}\)  increasing with growth temperature, while \(\beta _{T}\) decrease with growth temperature. At low temperatures, \(\beta _{T}\)  is larger than \(\beta _{L}\) ; while at high temperature, \(\beta _{T}\)  and \(\beta _{L}\) have comparable values. Although previously results have shown that  \(\beta _{L}\) can be either larger[3] or smaller than \(\beta _{T}\) [2], both the  \(\beta _{T}\) and \(\beta _{L}\)  would increase with growth temperature, our results are different with them, its physical origin is still needed to be further confirmed. Currently, the origin of the bulk SOT in L10-FePt is still in a debate, and recently Zhu. et. al suggested that the non-uniform strain plays a critical role in determining the SOT in L10-FePt.[4] The vicinal substrates, where the lattice stain could be continuously tuned by controlling by the vicinal angle, provides a unique way to investigate the dependence between the lattice stain and SOT. We characterize the \(\beta _{L(T)}\) for films grown on different vicinal substrates with \(\alpha =0^{\circ},5^{\circ},7^{\circ},10^{\circ}\) , as shown in Figure. 4g. We found both   \(\beta _{T}\)  and  \(\beta _{L}\)   do not significantly change with \(\alpha \) . Our results demonstrates that the lattice strain does not significantly influence the SOT. 
\section{ Conclusion}
In summary, our results present an artificial way to realize titled perpendicular anisotropy in L10-FePt film, which can be used to realize the field-free switching of the L10-FePt film. The switching can withstand a strong field disturbance (~1kOe), which is much larger than most of other established schemes. The dependence on the vicinal angle, film thickness, and growth temperature demonstrated the operation window for field-free switching of L10-FePt is wide. Our results extending beyond the established schemes to realize field-free switching, and not limit to L10-FePt, potentially it could be applied to other magnetic (L10-CoPt, L10-CoPd, L10-FePd et. al) and even the antiferromagnetic systems (L10-PtMn, L10-PdMn, et. al), as it is established that the vicinal substrates could tune the magnetic anisotropy of various magnetic and antiferromagnetic materials.[23, 24,30] The method we present here also process several unique advantages. For examples, compared with the established schemes by adding additional functional layers,[10,12,14] which would unavoidably complex the device’s structure and operation, which not only result in reduced stability and reliability, but also may cause higher power consumption due to the shunting effect.[9] Our strategy could avoid these disadvantages as no additional layer is needed. In addition, compared with the field-free switching scheme by growing wedge-shaped film structures, which could only be fabricated in small samples,[11] our strategy is more suitable for industrial production, as the vicinal substrates could be fabricated in wafer scale. Physically, our results also found the SOT in L10-FePt do not significantly influence by the vicinal substrates, which shed light on the recent debate of the origin of the “bulk” SOT in the L10-FePt film. 

\section{Acknowledgement}
We thank Dr. Zhang Jian, Ji Lianze, Zhang xuefeng, Huang shuai for help of lithography, and XRD characterization and help discussion. This work is supported by the “Pioneer” and “Leading Goose” R\(\&\)D Program of Zhejiang Province under (Grant No. 2022C01053), the National Natural Science Foundation of China (Grants No. 11874135), Key Research and Development Program of Zhejiang Province (Grants No. 2021C01039), and Natural Science Foundation of Zhejiang Province, China (Grants No. LQ20F040005 and No. LQ21A050001). 

\section{References:}
1.	Sun;  Murray;  Weller;  Folks; Moser, Monodisperse FePt nanoparticles and ferromagnetic FePt nanocrystal superlattices. Science (New York, N.Y.) 2000, 287 (5460), 1989-92.
2.	Tang, M.;  Shen, K.;  Xu, S.;  Yang, H.;  Hu, S.;  Lu, W.;  Li, C.;  Li, M.;  Yuan, Z.;  Pennycook, S. J.;  Xia, K.;  Manchon, A.;  Zhou, S.; Qiu, X., Bulk Spin Torque-Driven Perpendicular Magnetization Switching in L1(0) FePt Single Layer. Adv. Mater. 2020, 32 (31).
3.	Liu, L.;  Yu, J.;  Gonzalez-Hernandez, R.;  Li, C.;  Deng, J.;  Lin, W.;  Zhou, C.;  Zhou, T.;  Zhou, J.;  Wang, H.;  Guo, R.;  Yoong, H. Y.;  Chow, G. M.;  Han, X.;  Dupe, B.;  Zelezny, J.;  Sinova, J.; Chen, J., Electrical switching of perpendicular magnetization in a single ferromagnetic layer. Physical Review B 2020, 101 (22).
4.	Zhu, L.;  Ralph, D. C.; Buhrman, R. A., Unveiling the Mechanism of Bulk Spin-Orbit Torques within Chemically Disordered FexPt1-x Single Layers. Advanced Functional Materials 2021, 31 (36).
5.	Zheng, S. Q.;  Meng, K. K.;  Liu, Q. B.;  Chen, J. K.;  Miao, J.;  Xu, X. G.; Jiang, Y., Disorder dependent spin-orbit torques in L1(0) FePt single layer. Applied Physics Letters 2020, 117 (24).
6.	Mihai Miron, I.;  Gaudin, G.;  Auffret, S.;  Rodmacq, B.;  Schuhl, A.;  Pizzini, S.;  Vogel, J.; Gambardella, P., Current-driven spin torque induced by the Rashba effect in a ferromagnetic metal layer. Nature Materials 2010, 9 (3), 230-234.
7.	Liu, L.;  Lee, O. J.;  Gudmundsen, T. J.;  Ralph, D. C.; Buhrman, R. A., Current-Induced Switching of Perpendicularly Magnetized Magnetic Layers Using Spin Torque from the Spin Hall Effect. Physical Review Letters 2012, 109 (9).
8.	Liu, L.;  Pai, C.-F.;  Li, Y.;  Tseng, H. W.;  Ralph, D. C.; Buhrman, R. A., Spin-Torque Switching with the Giant Spin Hall Effect of Tantalum. Science 2012, 336 (6081), 555-558.
9.	Shao, Q.;  Li, P.;  Liu, L.;  Yang, H.;  Fukami, S.;  Razavi, A.;  Wu, H.;  Wang, K.;  Freimuth, F.;  Mokrousov, Y.;  Stiles, M. D.;  Emori, S.;  Hoffmann, A.;  Akerman, J.;  Roy, K.;  Wang, J.-P.;  Yang, S.-H.;  Garello, K.; Zhang, W., Roadmap of Spin-Orbit Torques. Ieee Transactions on Magnetics 2021, 57 (7).
10.	Fukami, S.;  Zhang, C.;  DuttaGupta, S.;  Kurenkov, A.; Ohno, H., Magnetization switching by spin-orbit torque in an antiferromagnet-ferromagnet bilayer system. Nature Materials 2016, 15 (5), 535-+.
11.	Yu, G.;  Upadhyaya, P.;  Fan, Y.;  Alzate, J. G.;  Jiang, W.;  Wong, K. L.;  Takei, S.;  Bender, S. A.;  Chang, L.-T.;  Jiang, Y.;  Lang, M.;  Tang, J.;  Wang, Y.;  Tserkovnyak, Y.;  Amiri, P. K.; Wang, K. L., Switching of perpendicular magnetization by spin-orbit torques in the absence of external magnetic fields. Nature Nanotechnology 2014, 9 (7), 548-554.
12.	Kong, W. J.;  Wan, C. H.;  Wang, X.;  Tao, B. S.;  Huang, L.;  Fang, C.;  Guo, C. Y.;  Guang, Y.;  Irfan, M.; Han, X. F., Spin-orbit torque switching in a T-type magnetic configuration with current orthogonal to easy axes. Nature Communications 2019, 10.
13.	Zhao, Z.;  Smith, A. K.;  Jamali, M.; Wang, J.-P., External-Field-Free Spin Hall Switching of Perpendicular Magnetic Nanopillar with a Dipole-Coupled Composite Structure. Advanced Electronic Materials 2020, 6 (5).
14.	Wu, H.;  Nance, J.;  Razavi, S. A.;  Lujan, D.;  Dai, B.;  Liu, Y.;  He, H.;  Cui, B.;  Wu, D.;  Wong, K.;  Sobotkiewich, K.;  Li, X.;  Carman, G. P.; Wang, K. L., Chiral Symmetry Breaking for Deterministic Switching of Perpendicular Magnetization by Spin-Orbit Torque. Nano Letters 2021, 21 (1), 515-521.
15.	Eason, K.;  Tan, S. G.;  Jalil, M. B. A.; Khoo, J. Y., Bistable perpendicular switching with in-plane spin polarization and without external fields. Physics Letters A 2013, 377 (37), 2403-2407.
16.	You, L.;  Lee, O.;  Bhowmik, D.;  Labanowski, D.;  Hong, J.;  Bokor, J.; Salahuddin, S., Switching of perpendicularly polarized nanomagnets with spin orbit torque without an external magnetic field by engineering a tilted anisotropy. Proceedings of the National Academy of Sciences of the United States of America 2015, 112 (33), 10310-10315.
17.	Liu, L.;  Qin, Q.;  Lin, W.;  Li, C.;  Xie, Q.;  He, S.;  Shu, X.;  Zhou, C.;  Lim, Z.;  Yu, J.;  Lu, W.;  Li, M.;  Yan, X.;  Pennycook, S. J.; Chen, J., Current-induced magnetization switching in all-oxide heterostructures. Nature Nanotechnology 2019, 14 (10), 939-+.
18.	Leroy, F.;  Mueller, P.;  Metois, J. J.; Pierre-Louis, O., Vicinal silicon surfaces: From step density wave to faceting. Physical Review B 2007, 76 (4).
19.	Tegenkamp, C., Vicinal surfaces for functional nanostructures. Journal of Physics-Condensed Matter 2009, 21 (1).
20.	Ma, S.;  Tan, A.;  Deng, J. X.;  Li, J.;  Zhang, Z. D.;  Hwang, C.; Qiu, Z. Q., Tailoring the magnetic anisotropy of Py/Ni bilayer films using well aligned atomic steps on Cu(001). Scientific Reports 2015, 5.
21.	Kawakami;  Escorcia, A.; Qiu, Symmetry-Induced Magnetic Anisotropy in Fe Films Grown on Stepped Ag(001). Physical review letters 1996, 77 (12), 2570-2573.
22.	Dhesi, S.;  Dürr, H.; Van der Laan, G., Canted spin structures in Ni films on stepped Cu (001). Physical Review B 1999, 59 (13), 8408.
23.	Choi, J.;  Wu, J.;  Wu, Y. Z.;  Won, C.;  Scholl, A.;  Doran, A.;  Owens, T.; Qiu, Z. Q., Effect of atomic steps on the interfacial interaction of FeMn/Co films grown on vicinal Cu(001). Physical Review B 2007, 76 (5).
24.	Zhu, J.;  Li, Q.;  Li, J. X.;  Ding, Z.;  Liang, J. H.;  Xiao, X.;  Luo, Y. M.;  Hua, C. Y.;  Lin, H. J.;  Pi, T. W.;  Hu, Z.;  Won, C.; Wu, Y. Z., Antiferromagnetic spin reorientation transition in epitaxial NiO/CoO/MgO(001) systems. Physical Review B 2014, 90 (5).
25.	He, P.;  Ma, L.;  Shi, Z.;  Guo, G.;  Zheng, J.-G.;  Xin, Y.; Zhou, S., Chemical Composition Tuning of the Anomalous Hall Effect in Isoelectronic L 1 0 FePd 1-x Pt x Alloy Films. Physical Review Letters 2012, 109 (6), 066402.
26.	Okamoto, S.;  Kikuchi, N.;  Kitakami, O.;  Miyazaki, T.;  Shimada, Y.; Fukamichi, K., Chemical-order-dependent magnetic anisotropy and exchange stiffness constant of FePt (001) epitaxial films. Physical Review B 2002, 66 (2), 024413.
27.	van den Brink, A.;  Vermijs, G.;  Solignac, A.;  Koo, J.;  Kohlhepp, J. T.;  Swagten, H. J. M.; Koopmans, B., Field-free magnetization reversal by spin-Hall effect and exchange bias. Nature Communications 2016, 7.
28.	Hayashi, M.;  Kim, J.;  Yamanouchi, M.; Ohno, H., Quantitative characterization of the spin-orbit torque using harmonic Hall voltage measurements. Physical Review B 2014, 89 (14).
29.	Garello, K.;  Miron, I. M.;  Avci, C. O.;  Freimuth, F.;  Mokrousov, Y.;  Bluegel, S.;  Auffret, S.;  Boulle, O.;  Gaudin, G.; Gambardella, P., Symmetry and magnitude of spin-orbit torques in ferromagnetic heterostructures. Nature Nanotechnology 2013, 8 (8), 587-593.
30.	Perna, P.;  Rodrigo, C.;  Jimenez, E.;  Mikuszeit, N.;  Teran, F. J.;  Mechin, L.;  Camarero, J.; Miranda, R., Magnetization reversal in half metallic La0.7Sr0.3MnO3 films grown onto vicinal surfaces. Journal of Applied Physics 2011, 109 (7).

\end{document}